\documentclass[aps,prl,twocolumn,notitlepage,groupedaddress,showpacs]{revtex4-1}

\usepackage{amsmath}    % need for subequations
\usepackage{amsfonts}
\usepackage{bm}
\usepackage{amssymb}
\usepackage{appendix}
\usepackage{graphicx}   % need for figures
\usepackage{color}      % use if color is used in text
\usepackage{subfigure}  % use for side-by-side figures
\usepackage{comment}
\begin{document}

\title{Complete photonic bandgaps in supercell photonic crystals}
\author{Alexander Cerjan and Shanhui Fan} 
%\email[]{???}
\affiliation{Department of Electrical Engineering, and Ginzton Laboratory, Stanford  University,  Stanford,  California  94305,  USA}

\date{\today}

\begin{abstract}
  We develop a class of supercell photonic crystals supporting complete photonic bandgaps based on breaking
  spatial symmetries of the underlying primitive photonic crystal. One member of this class based on a
  two-dimensional honeycomb structure supports a complete bandgap for an index-contrast ratio as low as $n_{high}/n_{low} = 2.1$,
  making this the first such 2D photonic crystal to support a complete bandgap in lossless materials at visible
  frequencies. The complete bandgaps found in such supercell photonic crystals do \textit{not} necessarily monotonically increase
  as the index-contrast in the system is increased, disproving a long-held conjecture of complete bandgaps in photonic
  crystals.
\end{abstract}

%\pacs{42.55.Ah, 42.55.Px, 42.70.Hj, 42.60.Pk}
\maketitle

Since their discovery, photonic crystals have become an indispensable technology across the entire
field of optical physics due to their ability to confine and control light of an arbitrary wavelength \cite{yablonovitch_inhibited_1987,john_strong_1987,joannopoulos_photonic_1997,joannopoulos}.
This critical feature is achieved by designing the crystal lattice to possess a complete photonic bandgap,
a range of frequencies for which no light can propagate regardless of its momentum or polarization.
Unlike their electronic counterparts in conventional crystals,
whose band structure is limited to the crystal lattices available in atomic and molecular structures,
the dielectric structure comprising a photonic crystal can be specified with nearly complete arbitrariness,
yielding a vast design space for optimizing photonic crystals for specific applications that is limited only
by the index of refraction of available materials at the operational wavelength.
For example, photonic crystals have been developed to
promote absorption in monolayer materials \cite{piper_total_2014,piper_crit_2014}, or for use in achieving high-power solid-state lasers \cite{noda_polarization_2001,kurosaka_-chip_2010,hirose_watt-class_2014}.
Moreover, this design freedom in dielectric structures has been leveraged in numerous studies to optimize the
complete bandgaps in high-index materials \cite{fan_design_1994,dobson_maximizing_1999,doosje_photonic_2000,cox_band_2000,johnson_three-dimensionally_2000,shen_large_2002,biswas_three-dimensional_2002,maldovan_photonic_2002,maldovan_exploring_2003,michielsen_photonic_2003,sigmund_systematic_2003,toader_photonic_2003,jensen_systematic_2004,stanley_inverse_2004,kao_maximizing_2005,maldovan_photonic_2005,halkjaer_maximizing_2006,watanabe_broadband_2006,sigmund_geometric_2008,men_bandgap_2010,jia_two-pattern_2011,liang_formulation_2013}. 
Unfortunately, similar efforts to realize new crystal structures or improve upon existing ones
to achieve complete bandgaps in low-index materials have
yielded only minor improvements upon traditional simple crystal structures with high symmetry \cite{oskooi_zerogroup-velocity_2009,men_robust_2014},
i.e.\ the inverse triangular lattice in two-dimensions \cite{joannopoulos} and the network diamond lattice in three-dimensions \cite{maldovan_diamond-structured_2004}.
This has led many to conclude that the known high-symmetry dielectric structures are nearly optimal for achieving
low-index complete bandgaps \cite{men_robust_2014}.

However, the ability to realize complete bandgaps for
low-index materials is critically important to the development of many photonics
technologies operating in the visible wavelength range, such as augmented and virtual reality systems, where the
highest index lossless materials have $n \approx 2.4$--$2.5$. Currently, there are no known 2D photonic crystals
which display a complete bandgap in this index contrast regime, and thus it is not possible to realize dual-polarization in-plane guiding
at this index contrast using photonic crystal slabs. Although a few 3D photonic crystals do display a complete bandgap in this range,
3D photonic crystals are difficult to fabricate \cite{yablonovitch_photonic_1993,lin_three-dimensional_1998,noda_full_2000,vlasov_-chip_2001,qi_three-dimensional_2004}.
%Additionally, realizing low-index complete photonic bandgaps
%would also increase the available materials for realizing slow-light devices \cite{oskooi_zerogroup-velocity_2009}.

In this Letter, we demonstrate a new class of complete photonic bandgaps which are achieved
by judiciously breaking symmetry, rather than promoting it. By starting with a photonic crystal possessing
a large bandgap for one polarization, we show that by expanding the primitive cell of the photonic crystal to form a supercell
and then slightly adjusting the dielectric structure within this supercell to break part of the translational symmetry of the original primitive cell,
a bandgap in the other polarization can be opened, thus producing a complete bandgap. This method
yields a two-dimensional photonic crystal based on a honeycomb lattice with a complete bandgap that persists down to an
index-contrast ratio of $n_{high}/n_{low} = 2.1$, the lowest known index-contrast ratio in 2D photonic crystals.
Such low index contrast bandgaps can also be translated into photonic crystal slabs, where they represent
the first structures able to confine optical frequencies in-plane regardless of their polarization.
In contrast to the complete photonic bandgaps found in traditional photonic crystals, complete bandgaps in supercell photonic crystals
do not necessarily monotonically increase as a function of the index-contrast ratio, disproving a long-held conjecture
in the photonic crystal literature \cite{joannopoulos}.

\begin{figure}[t!]
  \centering
  \includegraphics[width=0.48\textwidth]{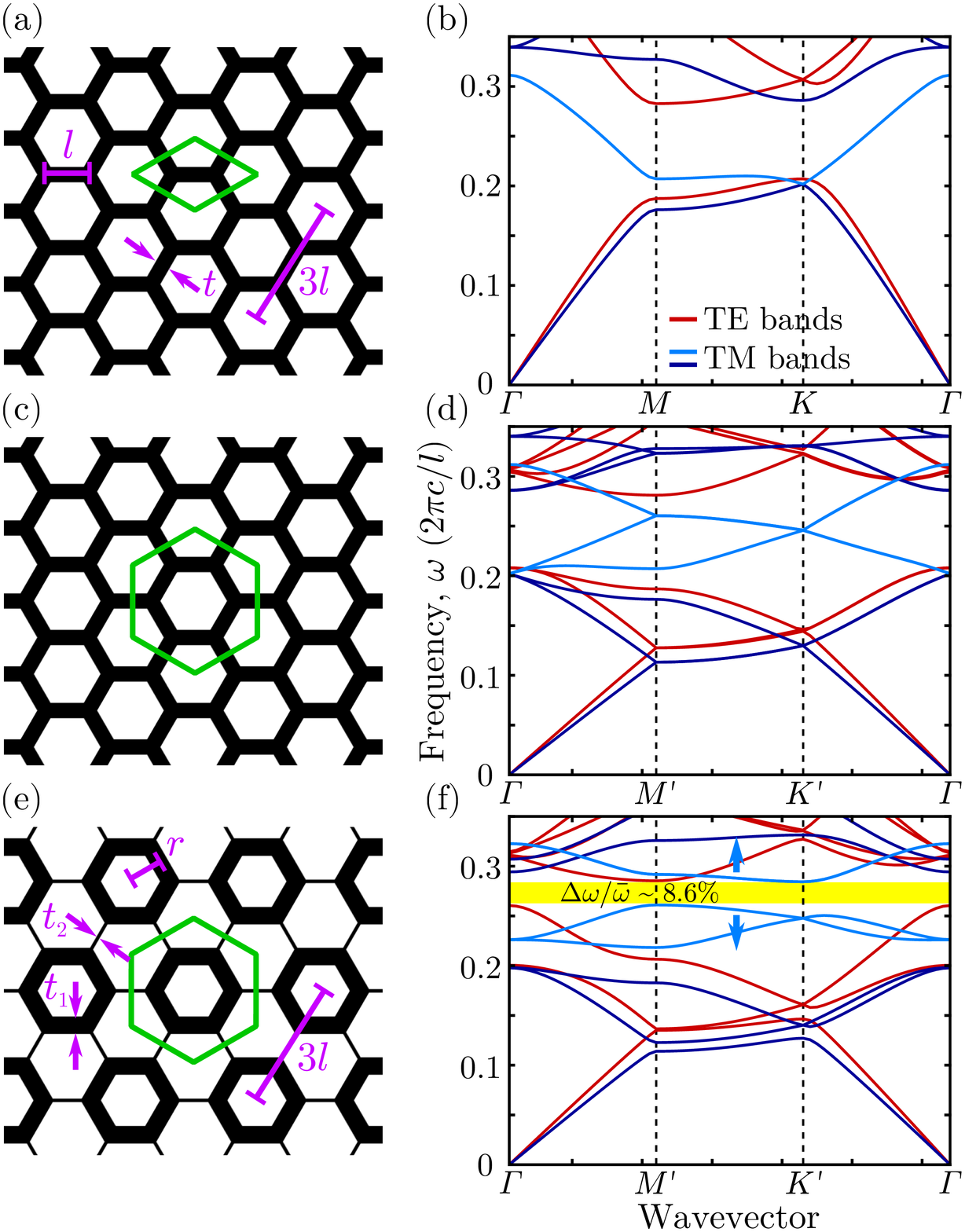}
  \caption{
    (a),(c),(e) Schematics of 2D photonic crystals formed of a network of a high dielectric
    material with $n=2.4$ (black region) in a background of air, $n_{\textrm{air}}=1$. The unit cell
    for the corresponding band structure in (b),(d),(f) is shown in green. Parameters which characterize
    these crystals are indicated in purple. The structures in (a) and (c) are completely specified
    by the line thickness, $t$, while the structure in (e) is completely specified by three parameters,
    the thick line thickness, $t_1$, the thin line thickness, $t_2$, and the shortest distance from a thick line
    to the center of its corresponding thick-bordered hexagon, $r$. $l$ denotes the distance between two
    vertices in the primitive honeycomb lattice. (b),(d),(f) Band structures around the border of the irreducible Brillouin Zone
    for the photonic crystals shown in (a),(c),(e), respectively, in which the TE bands are shown
    in red and the TM bands are shown in blue. In (b), the 2$^{\textrm{nd}}$ TM band is shown as
    cyan, while the corresponding folded bands in the supercell Brillouin zone in (d) and (f)
    are also shown in cyan.
    In (b),(d), $t/l=0.3636$, while in (f), $t_1/l = 0.4149$, $t_2/l = 0.0816$, and $r/l = 0.8145$,
    and a complete bandgap is found with width $\Delta \omega / \omega_0 = 8.6\% $ between the
    8$^{\textrm{th}}$ and 9$^{\textrm{th}}$ bands, shown in yellow.
    Band structures were calculated using \textsc{MIT Photonic Bands} (MPB) \cite{MPB}.
    \label{fig:1}}
\end{figure}

To illuminate how symmetry breaking can help to realize complete photonic bandgaps, we first consider
the 2D photonic crystal comprised of a network structure on a honeycomb lattice depicted in Fig.\ \ref{fig:1}(a).
The primitive cell of this system contains a pair of vertices in this network lattice, and the system can
be parameterized solely in terms of the thickness, $t$, of the lines forming the network structure.
Although in a low-index network structure, $n_{high}/n_{low} = 2.4$, a wide range of $t$ yields a large transverse electric (TE) bandgap as shown
in Fig.\ \ref{fig:1}(b), no complete photonic bandgap exists for any choice of $t$ for this choice of $n_{high}/n_{low}$.

However, starting from the crystal structure as shown in Fig.\ \ref{fig:1}(a), we can find
a complete bandgap in a closely related supercell photonic crystal. First, we increase
the size of the primitive cell to contain six vertices which form the supercell, as depicted in
Fig.\ \ref{fig:1}(c). In doing so, each of the bands in the primitive Brillouin zone fold up into
three bands in the supercell Brillouin zone, shown in Fig.\ \ref{fig:1}(d). %In particular, note that
Along the edge of the supercell Brillouin zone ($M' \rightarrow K'$), pairs of the folded supercell
bands can form lines of degeneracies, i.e.\ degenerate contours \cite{cerjan_zipping_2016}, and one
such degenerate contour is formed per trio of folded bands originating from the same band in the primitive Brillouin zone.
From the perspective of the
supercell photonic crystal, the degeneracies comprising each of the degenerate contours are accidental,
and are only the result of the supercell obeying an extra set of spatial symmetries as it is a three-fold
copy of the original photonic crystal. Thus by breaking these symmetries, the degeneracies forming
the degenerate contours are lifted, and a gap can begin to open between the two transverse magnetic (TM) bands.
The supercell is now characterized in terms of three parameters, the thickness
of the center lines, $t_1$, the thickness of the connecting lines, $t_2$, and the size of the
thick-lined hexagons, $r$, shown in Fig.\ \ref{fig:1}(e). When the symmetry breaking becomes sufficiently strong,
a complete photonic bandgap opens between the $8$th and $9$th bands of the system, whose maximum width at $n_{high}/n_{low} = 2.4$ can be
found numerically to be $\Delta \omega / \bar{\omega} = 8.6\% $.
Here, $\Delta \omega$ is the difference between the minimum of the $9$th band and the maximum of the $8$th
band, while $\bar{\omega}$ is the central frequency between the two bands.
Rigorously, after the symmetry of this system is broken, the supercell containing six vertices becomes
the primitive cell of the system. However, for semantic convenience, we will continue to refer to
this larger primitive cell as the `supercell' and reserve `primitive cell' for the smaller system
whose symmetry is intact.
%This example also illustrates another important feature of bandgaps in supercell photonic crystals,
%the bandgap is found between higher order bands.

\begin{figure}[t!]
  \centering
  \includegraphics[width=0.48\textwidth]{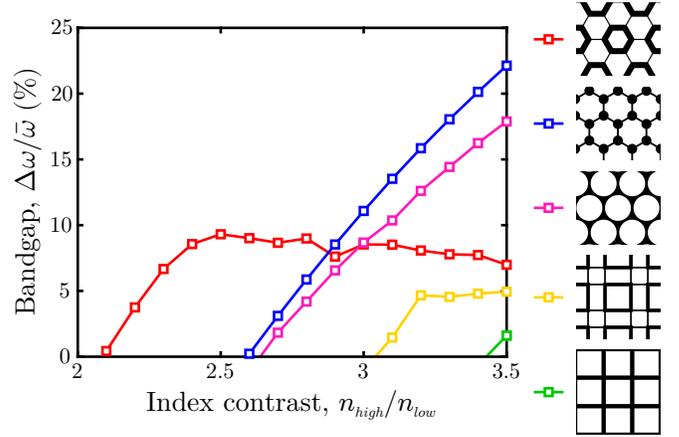}
  \caption{
    Plot of the optimized complete bandgap width as a function of the index contrast, $n_{high}/n_{low}$
    for five different 2D photonic crystals,
    supercell network honeycomb lattice discussed in Fig.\ \ref{fig:1} (red), traditional inverse triangular lattice (magenta),
    decorated network honeycomb lattice designed in Ref. \cite{oskooi_zerogroup-velocity_2009} (blue), traditional network square lattice (green), and
    supercell network square lattice discussed in Fig.\ \ref{fig:square} (yellow).
    The black regions correspond to the high index material, $n_{high}$, while the white
    regions correspond to low index material, $n_{low}$.
    Optimized complete bandgaps were calculated using MPB \cite{MPB}.
    \label{fig:master}}
\end{figure}

Previously, the lowest index-contrast ratio known to support a complete bandgap in a 2D photonic
crystal was a decorated honeycomb lattice, shown as the blue curve in Fig.\ \ref{fig:master}, which has a complete
bandgap between the $3$rd and $4$th bands for index-contrast ratios as low as $n_{high}/n_{low} = 2.6$ \cite{oskooi_zerogroup-velocity_2009}.
This structure provides a relatively modest improvement upon the traditional triangular lattice of air holes, also shown
in Fig.\ \ref{fig:master} as the pink curve.
In contrast, the supercell honeycomb lattice possesses a complete bandgap for index-contrast ratios
as low as $n_{high}/n_{low} = 2.1$, and as such is the first 2D photonic crystal design that can
realize a complete bandgap for visible wavelengths where the largest index of refraction possible
in lossless materials is $n \approx 2.4$--$2.5$, which is found in Diamond \cite{phillip_kramers-kronig_1964}, Titanium Dioxide \cite{devore_refractive_1951}, and Strontium Titanate \cite{weber_crc_1994}.
Furthermore, this supercell honeycomb structure could also be used in conjunction with high-index materials
available in other frequency ranges so that the low-index material used in the structure need not be air. For example,
this could enable realizing complete bandgaps in completely solid photonic crystal fibers
operating in the communications band, where the high index regions are Silicon, $n_{high} = 3.48$, and the
low-index regions are filled with fused silica, $n_{low}=1.45$, such that $n_{high}/n_{low} = 2.4$ for $\lambda = 1.55\mu$m.

\begin{figure}[t!]
  \centering
  \includegraphics[width=0.48\textwidth]{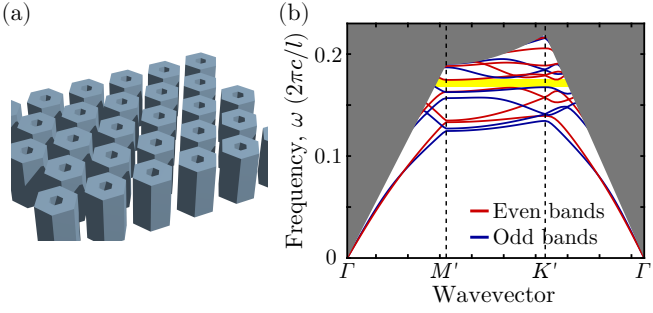}
  \caption{
    (a) Schematics of a 3D photonic crystal slab consisting of high dielectric hexagonal rods
    with $n=2.4$ in a background of air, $n_{\textrm{air}}=1$, each of which also contains a hexagonal air hole
    in its center. This system can be completely parameterized using exactly the same choice
    of $t_1$, $t_2$, and $r$ from the 2D photonic crystal shown in Fig.\ \ref{fig:1}(e), as
    well as the height of each rod, $h$. For the structure in (a) $t_2 = 0$. (b) Band diagram for the photonic crystal slab shown
    in (a), in which the even modes with respect to the mirror plane at $z=0$ are shown in
    red, while the odd modes are shown in blue. The grey regions of the band diagram indicate
    the continuum of radiation modes which lies above the light line. As can be seen, a complete
    bandgap with width $\Delta \omega / \omega_0 = 4.3\% $ (yellow) opens between the 7$^{\textrm{th}}$ and 8$^{\textrm{th}}$ bands for
    $t_1/l = 0.7482$, $t_2/l = 0$, $r/l = 0.7405$, and $h/l = 3.564$, where $l$ is again the distance
    between two vertices in the underlying primitive honeycomb lattice.
    Band structures were calculated using MPB \cite{MPB}.
    \label{fig:slab}}
\end{figure}

\begin{figure}[t!]
  \centering
  \includegraphics[width=0.48\textwidth]{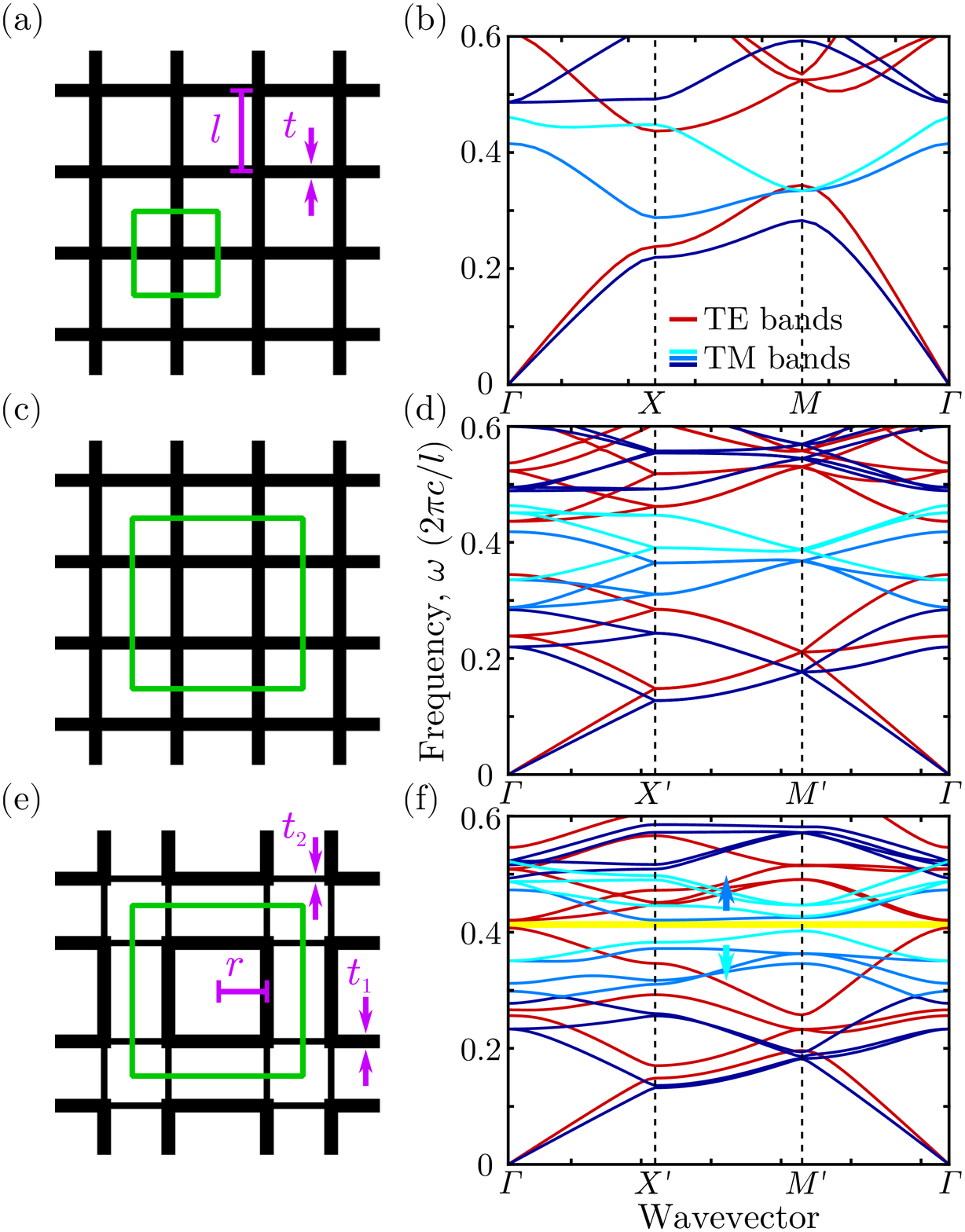}
  \caption{
    (a),(c),(e) Schematics of 2D photonic crystals formed of a network of a high dielectric
    material with $n=3.2$ (black region) in a background of air, $n_{\textrm{air}}=1$. The unit cell
    for the corresponding band structure in (b),(d),(f) is shown in green. Parameters which characterize
    these crystals are indicated in purple. The structures in (a) and (c) are completely specified
    by the line thickness, $t$, while the structure in (e) is completely specified by three parameters,
    the thick-line thickness, $t_1$, the thin-line thickness, $t_2$, and the shortest distance from a thick line
    to the center of its corresponding thick-line bordered square, $r$. $l$ denotes the distance
    between two vertices in the primitive cell. (b),(d),(f) Band structures around the border of the irreducible Brillouin Zone
    for the photonic crystals shown in (a),(c),(e), respectively, in which the TE bands are shown
    in red and the TM bands are shown in blue. In (b), the 2$^{\textrm{nd}}$ (3$^{\textrm{rd}}$) TM band is shown as
    light blue (cyan), while the corresponding folded bands in the supercell Brillouin zone in (d) and (f)
    are also shown in light blue (cyan).
    In (b),(d), $t/a_{\textrm{PC}}=0.32$, while in (f), $t_1/a_{\textrm{PC}} = 0.34$, $t_2/a_{\textrm{PC}} = 0.16$, and $r/a_{\textrm{PC}} = 0.605$,
    and a complete bandgap is found with width $\Delta \omega / \omega_0 = 4.6\% $ between the
    12$^{\textrm{th}}$ and 13$^{\textrm{th}}$ bands, shown in yellow.
    Band structures were calculated using MPB \cite{MPB}.
    \label{fig:square}}
\end{figure}

This 2D supercell honeycomb structure can also be used to design new photonic crystal slabs
so as to provide confinement in three dimensions. In Fig.\ \ref{fig:slab}, we show a supercell
honeycomb slab with a complete below-light-line dual-polarization bandgap of $\Delta \omega / \omega_0 = 4.3\% $ for $n_{high}/n_{low} = 2.4$.
Note that in photonic crystal slabs, to define a  band gap one only considers the phase space regions below the light line,
as above the light line the radiations modes form a continuum with no gaps.
To the best of our knowledge, this represents the
first system which could confine visible frequencies emitted from a omni-polarization source burried within the system.
Likewise, this design could also be used to realize entirely solid photonic crystal slabs for communications
frequencies where higher index dielectric materials are available.

The procedure used above is not restricted to the honeycomb lattice.
To illustrate this point, we use the same method to produce a
complete bandgap in a 2D supercell square network lattice, as shown in Fig.\ \ref{fig:square}. Unlike
in the primitive honeycomb crystal,
the TE bandgap in the underlying primitive square lattice is spanned by two TM bands.
Thus, a complete bandgap is only realized for sufficiently strong symmetry breaking so that
not only does a gap open in each degenerate contour of the folded supercell TM bands, but that
a gap opens between these two folded bands. This limits the overall width of the complete bandgap,
and the lowest index-contrast ratio for which this structure possesses a complete bandgap is $n_{high}/n_{low} \sim 3.1$,
as shown as the yellow line in Fig.\ \ref{fig:master}. However, this still represents a significant improvement
upon the range of index-contrast ratios which can yield a complete bandgap when compared against
other 2D photonic crystals based on a square lattice.

Complete bandgaps in supercell photonic crystal possess three features which distinguish them from complete
bandgaps found in traditional photonic crystals. First, as noted above, these structures have been designed
by specifically breaking symmetry within the system. This is entirely distinct from what is observed for
bandgaps found in traditionally designed structures, which consider the high-symmetry triangular
lattice in 2D and diamond lattice in 3D to be near optimal. 
Second, as can be seen in Fig.\ \ref{fig:master}, complete bandgaps in supercell structures do not necessarily
monotonically increase in size as a function of the index-contrast. This disproves a long-held conjecture of
complete bandgaps in photonic crystals, that the optimized bandgap (between the same two bands) always increases
as the index contrast increases \cite{joannopoulos}.
Finally, the complete bandgap in supercell crystals is found between higher order bands.
This is unlike many photonic crystal structures previously considered where the complete bandgap
occurs between lower-order bands. 

%At this juncture, it is natural to ask that given the extensive literature on optimizing
%photonic crystal bandgaps stretching back two decades \cite{fan_design_1994,dobson_maximizing_1999,doosje_photonic_2000,cox_band_2000,johnson_three-dimensionally_2000,shen_large_2002,biswas_three-dimensional_2002,maldovan_photonic_2002,maldovan_exploring_2003,michielsen_photonic_2003,sigmund_systematic_2003,toader_photonic_2003,jensen_systematic_2004,stanley_inverse_2004,kao_maximizing_2005,maldovan_photonic_2005,halkjaer_maximizing_2006,watanabe_broadband_2006,sigmund_geometric_2008,men_bandgap_2010,jia_two-pattern_2011,liang_formulation_2013,oskooi_zerogroup-velocity_2009,men_robust_2014},
%how have these supercell structures eluded discovery?

Designing two-dimensional supercell photonic crystals to possess complete bandgaps has three steps. First,
a candidate primitive photonic crystal must be constructed which possesses a large bandgap for one
polarization, and which is spanned by at most one or two bands in the other polarization. Second,
a supercell must be generated from this primitive cell such that the degenerate contours of the
folded band spanning the single-polarization bandgap lie entirely within the single-polarization bandgap.
Finally, the supercell perturbation which breaks the underlying primitive cell symmetries must be designed,
such that a bandgap in the degenerate contour opens before the single-polarization bandgap in the original
primitive system closes.

We expect these same design principles to hold for finding complete bandgaps in three-dimensional
supercell photonic crystals, but in practice we have been unable to find such a structure.
Although the second and third steps in the above procedure are relatively straightforward, finding
good candidate primitive cell structures is much more challenging in 3D, as it is rare to find
what would be a large bandgap spanned by only a single other band. For comparison, this is relatively easy in
2D, structures with isolated dielectric elements typically possess large TM bandgaps, but not TE bandgaps, while network
structures typically possess large TE bandgaps, but no TM bandgaps.

In conclusion, we have developed a new class of photonic crystals which support complete bandgaps
which stem from breaking spatial symmetries. These structures can exhibit complete bandgaps
for much lower index-contrast ratios than was previously known, enabling the confinement of
visible light in two-dimensional structures. The discovery of this new class of supercell structures also
provides encouragement that there may be significant improvements remaining to be discovered in
designing and optimizing complete bandgaps at low index-contrasts in both two- and three-dimensional systems.

\begin{acknowledgments}
This work was supported by an AFOSR MURI program (Grant No.\ FA9550-12-1-0471), and an AFOSR project
(Grant No.\ FA9550-16-1-0010).
\end{acknowledgments}

%\bibliography{pbg_references}


\end{document}